\documentclass[pre,twocolumn,aps,floats,longbibliography,
floatfix,showpacs,showkeys,amsmath,amssymb,nofootinbib]{revtex4-1}

\usepackage{dcolumn}% Align table columns on decimal point
\usepackage{bm}% bold math
\usepackage{txfonts}% bold math
\usepackage{pxfonts}% bold math
\usepackage{graphicx}
\usepackage{psfrag}
\usepackage{times}
\usepackage{multirow} 

\begin{document}

\title{Logistic forecasting of GDP competitiveness}

%\author{Abhin Kakkad}\email{p19abhink@iima.ac.in; abhinkakkad@gmail.com}
%\affiliation{Indian Institute of Management Ahmedabad, 
%Vastrapur, Ahmedabad 380015, Gujarat, India}

%\author{Harsh Vasoya}
%\email{harshvasoya008@gmail.com}
%\affiliation{Dhirubhai Ambani Institute of Information 
%\& Communication Technology, Gandhinagar, Gujarat 382007, India\\ }

\author{Arnab K. Ray}\email{arnab\_kumar@daiict.ac.in; arnab.kr311@gmail.com}
\affiliation{Dhirubhai Ambani Institute of Information and 
Communication Technology, Gandhinagar 382007, Gujarat, India}

\date{\today}

\begin{abstract}
The GDP growth of national economies is modelled by the logistic 
function. 
Applying it on the GDP data of the World Bank till the year 2020, 
we forecast the outcome of the competitive GDP growth of Japan, 
Germany, UK and India, all of whose current GDPs are very close to one 
another. Fulfilling one of the predictions, in 2022 the GDP of India 
has indeed overtaken the GDP of UK. Our overall forecast is that 
by 2047, the GDP of India will be greater than that of the other 
three countries. We argue that when trade saturates, large and 
populous countries (like India) have the benefit of high domestic 
consumption to propel their GDP growth. 
\end{abstract}

%\pacs{89.65.Gh, 05.40.Jc, 05.40.Fb, 05.45.Tp}
%\keywords{Econophysics, financial markets; Brownian motion; 
%Random walks; Time series analysis}

\maketitle

\section{Introduction}
\label{sec1} 
The logistic equation is a standard example of a first-order
autonomous nonlinear dynamical system~\citep{stro}. 
Introduced originally to study population dynamics~\citep{stro,braun}, 
it was later applied to multiple problems of
socio-economic~\citep{braun,mon78,akr10,kr22} and scientific
interest~\citep{stro}.
%This arouses curiosity about the general import
%of the logistic equation. 
This is because the growth of many natural systems
is modelled quite accurately by the logistic equation,
the growth of species being one of many such examples~\citep{braun}.
Hence, the logistic equation is organically compatible with
natural evolution in a free and productive environment.
This principle can be extended to the evolution of
economic systems as well, a point of view that is supported 
by the generally successful logistic modelling of the GDP and 
trade dynamics of some leading national economies~\citep{kr22}. 
 
The GDP (an abbreviation of Gross Domestic Product) of a country
is the market value of goods and services produced by the country 
in a year~\citep{samnord,mas07,gmacl07}. GDP thus quantifies the 
aggregate outcome of the economic activities of a country that are 
performed all round the year. As such, the GDP of a national economy 
is a dynamic quantity and its evolution (commonly implying growth) 
can be followed through time. %, scaled in years. 
To this very end, the logistic equation turns out to be a simple 
and convenient mathematical tool, as has been shown in an earlier
study carried out on countries that are ranked high globally in terms 
of their national GDPs~\citep{kr22}. 

From a macroeconomic perspective, GDP
is a standard yardstick with which the state of a national economy
is gauged, and in a global comparison of national economies, 
the GDP of a country is a reliable point of reference. By this
criterion, globally the top six economies 
pertain to USA, China, Japan, 
Germany, UK and India. At present %(2021) 
these six countries account 
for nearly 60\% of the global GDP and nearly 40\% of the global trade. 
China, India and USA are the three most populous countries in the
world, accounting for almost 40\% of the world population. On the 
scale of strategic economic regions, 
the three most dominant economies in the North-Atlantic region are
USA, Germany and UK. Likewise, the three most dominant economies in 
the Indo-Pacific region are China, Japan and India.
All six 
countries are members of important economic blocs like G7 and BRICS. 
USA, Japan, Germany and UK belong to the former bloc, while China 
and India belong to the latter. 
Besides, all of these countries are the leading global
representatives of three types of economic systems, 
namely, free economies (USA, Japan, Germany and UK), controlled 
economies (China) and mixed economies (India). 
That only six countries should exert such an
overarching influence on the global economy is compatible with the 
scale-free (power law) degree distribution of GDP~\citep{gl04,gmacl07},  
%These features would not be qualitatively altered if more countries 
%were to be included in our survey. 
%For example, G20, which is an 
%economic bloc comprising the European Union and nineteen 
%independent countries (including the six that we consider here),
%accounts for 80\% of the global GDP, 75\% of the global 
%trade and 60\% of the world population. 
because in 
scale-free distributions the disproportionate
dominance of a few elements is a natural occurrence~\citep{albar02}. 
To this the global economic order 
can be no exception. Summing up these facts, we can now  
argue that our study on a restricted scale of six countries  
(which are global leaders in terms of their GDPs) adequately  
represents the essence of the GDP growth of more countries 
that can be studied on a larger scale.  
%the general essence of GDP growth 

Country-wise annual GDP data, on which we have based our modelling 
and analysis, have been collected from the World 
Bank website~\citep{usgd,cngd,jpgd,degd,ukgd,ingd} up to the year 2020. 
%The initial year of the GDP data for USA, China, Japan, UK and 
%India is 1960. For Germany, the data  begin from 
%1970. All data sets end either in 2019 or 2020. Hence, our study 
%spans across six decades in all cases but one. 
%For all the six countries, GDP is universally measured
%in terms of US dollars.
The basic mathematical theory of the logistic equation and its 
application on the GDP data are laid out in Sec.~\ref{sec2}. 
The numerical and statistical analyses of the modelling 
are summarized in Table~\ref{t1}.
In Sec.~\ref{sec3} we consider the competitive GDP growth of 
Japan, Germany, UK and India. Extrapolating the theoretical logistic 
functions (all calibrated by the GDP data~\citep{jpgd,degd,ukgd,ingd}) 
beyond 2020, we predict the specific years in which 
the GDP of one country will overtake the GDP of another. 
Three such overtakes are to occur in the future, one of which has
already happened in 2022, in precise agreement with our forecast. 
In Sec.~\ref{sec4} we remark on the impact of  
current geopolitics and adverse climatic events on GDP competitiveness.  

\section{Logistic modelling of GDP growth}
\label{sec2} 
The GDPs of all the six countries in this study are measured 
in US dollars. Quantifying GDP by the variable $G(t)$, in 
which $t$ is time (measured in years), we set up a first-order
autonomous dynamical system for $G(t)$ as~\citep{stro,kr22}  
\begin{equation}
\label{gdplog} 
\dot{G} \equiv \frac{{\mathrm d}G}{{\mathrm d}t} %%{\mathcal G}(G)
= \gamma G\left(1 - \frac{G}{k}\right). 
\end{equation}
Eq.~(\ref{gdplog}) is the well-known logistic equation. Its integral 
solution, under the initial condition of $G(0)=G_0$, is 
\begin{equation} 
\label{sologis} 
G(t) = \frac{G_0 e^{\gamma t}}{1+\left(G_0/k \right )
\left(e^{\gamma t}-1\right)}, 
\end{equation}
which is the logistic function. 
The time scale that is implicit in Eq.~(\ref{sologis}) 
is $\gamma^{-1}$. 
On early time scales, when $t \ll \gamma^{-1}$, the growth of $G$ 
in Eq.~(\ref{sologis}) can be  
approximated to be exponential, i.e. $G \simeq G_0 \exp(\gamma t)$.
This gives $\ln G \sim \gamma t$, which is a linear relation on a 
linear-log plot. We interpret $\gamma \simeq \dot{G}/G$
as the relative % (or fractional) 
growth rate in the early exponential regime. However, this exponential 
growth is not indefinite, and on time scales of 
$t \gg \gamma^{-1}$ (or $t \longrightarrow \infty$)
there is a convergence to $G=k$. Thus, according to Eq.~(\ref{sologis}), 
growth saturates to a finite limit on long time scales.
The transition from the exponential regime to the saturation 
regime occurs when $t \sim \gamma^{-1}$. 

\begin{figure}[]
\begin{center}
\includegraphics[scale=0.65, angle=0]{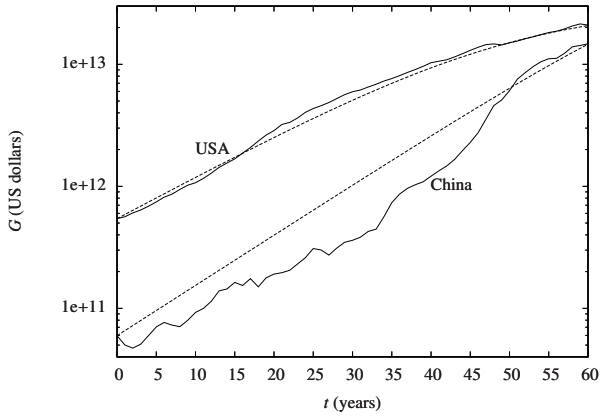}
\caption{\label{f1}\small{
Comparing the GDP growth of USA and China, which are, respectively,
the countries with the highest and the second highest GDPs in the world.
The smooth dotted curves model the GDP growth of both countries
according to Eq.~(\ref{sologis}), with the values of $\gamma$ and
$k$ in Table~\ref{t1}.
The World Bank data of the annual GDP
from 1960 ($t=0$) to 2020~\citep{usgd,cngd} show a much more ordered
progression for
USA than for China. Consequently, the logistic modelling of the GDP
growth of USA shows a greater closeness with the actual data than what
it does in the case of China.
}}
\end{center}
\end{figure}
\begin{figure}[]
\begin{center}
\includegraphics[scale=0.65, angle=0]{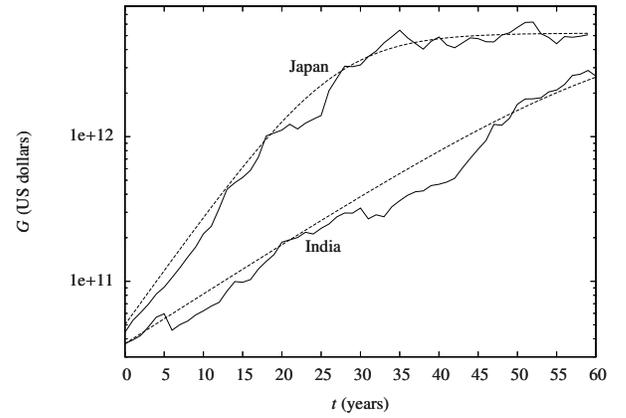}
\caption{\label{f2}\small{
Comparing the GDP growth of Japan and India, which, after China, 
are, respectively,
the countries with the second and the third highest GDPs in the
Indo-Pacific region.
The World Bank data of the annual GDPs of both 
countries start from 1960 ($t=0$)~\citep{jpgd,ingd}. 
The GDP data end in the year 2019 for 
Japan, and the year 2020 for India. The GDP growth of Japan has
a steep gradient in the early years, but by the year 2000, the 
growth has visibly stagnated. Both of these features are 
modelled closely by the logistic function --- the smooth dotted curve.  
In contrast, the GDP growth of India has been slow but on the 
whole steady, and by the year 2020, the GDP of 
India grows with a higher gradient than the GDP of Japan. At
this rate, the GDP of India will eventually 
overtake the GDP of Japan. This forecast is theoretically modelled 
in Fig.~\ref{f4} by extrapolating the  
logistic curves of Japan and India beyond 2020. 
These two theoretical logistic curves model the GDP growth of both 
countries
according to Eq.~(\ref{sologis}), with the values of $\gamma$ and
$k$ in Table~\ref{t1}. 
}}
\end{center}
\end{figure}
\begin{figure}[]
\begin{center}
\includegraphics[scale=0.65, angle=0]{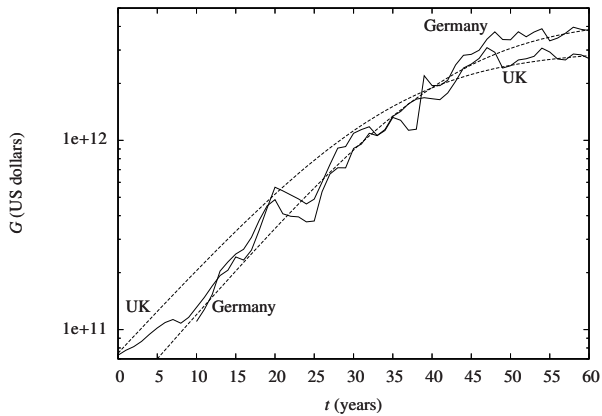}
\caption{\label{f3}\small{
Comparing the GDP growth of Germany and UK, which, after USA, are, 
respectively,
the countries with the second and the third highest GDPs in the 
North-Atlantic region. 
The World Bank data of the annual GDP of UK start from 
1960 ($t=0$) and end at 2020~\citep{ukgd}. For Germany, however, the 
GDP data~\citep{degd} start from 
1970 ($t =10\,\mathrm{years}$)~\citep{degd}. 
Till 1999-2000, both countries  
ran each other very close in terms of their GDP growth. Thereafter, 
the GDP of Germany has continuously led the GDP of UK. The 
beginning of the lead for Germany is theoretically
captured by the intersection of the smooth dotted curves 
around the year 2000 (shown clearly in Fig.~\ref{f4}). 
These two theoretical logistic curves model the GDP growth of 
both countries
according to Eq.~(\ref{sologis}), with the values of $\gamma$ and
$k$ in Table~\ref{t1}. In the case of Germany the 
theoretical logistic curve has been extrapolated backward before 1970. 
}}
\end{center}
\end{figure}

Of the six countries in our study, 
the initial year of the GDP data for USA, China, Japan, UK and 
India is 1960. For Germany, the data  begin from 
1970. All data sets end either in 2019 or 2020. Hence, our study 
spans across six decades in all cases but one. 
%For all the six countries, GDP is universally measured
%in terms of US dollars.
USA, Germany and UK are the top three 
economies in the North-Atlantic region, and China, Japan and India 
are likewise in the Indo-Pacific region. Moreover, USA and China are 
the top two economies of the world, with their respective GDPs being 
of the order of 20 trillion US dollars. The GDPs of Japan, Germany, 
UK and India are each approximately a quarter of the GDPs of 
either USA or China. 
Hence, on a scale of global competitiveness, we exclusively compare 
the GDP growth of USA and China in Fig.~\ref{f1}. After China, the two 
competing economies in the Indo-Pacific region are Japan and India. 
Their GDP growth is compared jointly in Fig.~\ref{f2}. Similarly, after
USA, the two competing economies in the North-Atlantic region are 
Germany and UK, whose GDP growth is compared together in Fig.~\ref{f3}.    
The early exponential growth of the GDP and its later convergence 
to a finite limit, as implied by Eq.~(\ref{sologis}), are modelled 
in all the linear-log plots in Figs.~\ref{f1},~\ref{f2}~and~\ref{f3}.  
The uneven lines follow the movement of the real GDP
data, available from the World Bank~\citep{usgd,cngd,jpgd,degd,ukgd,ingd}.
The smooth dotted curves theoretically model the real data with 
%the integral solution of Eq.~(\ref{gdplog}), which will be in the form of 
Eq.~(\ref{sologis}).
The values of $\gamma$ (the relative annual growth rate of GDP in 
the early stage) and
$k$ (the predicted maximum value of GDP), 
calibrated through the model fitting in all the cases, are to be
found in Table~\ref{t1}. The most
convincing match of the GDP data with the logistic function is seen
in Fig.~\ref{f1}, for USA. Consistent fitting of the GDP data
with the logistic function is also seen for Japan, Germany, UK and India,
for which Figs.~\ref{f2}~and~\ref{f3} provide evidence.
Similar consistency, however, is not observed in the model fitting
of the GDP data for China, as we note from the lower plot
in Fig.~\ref{f1}. These observations about the model-fitting of the
GDP data are statistically summarized in Table~\ref{t1}, which sets
down the mean $\mu$ and the standard deviation $\sigma$ of
the yearly relative variations of the actual GDP
data~\citep{usgd,cngd,jpgd,degd,ukgd,ingd} about the
theoretical logistic function.
Going by the values in Table~\ref{t1}, we contend that the 
natural and 
balanced growth of the GDP of a country can be gauged from the 
closeness between the theoretical logistic function and the actual 
GDP data. In support of this view, the GDP growth of USA is 
a compelling example. Conditions that favour such a GDP growth are 
discussed in Sec.~\ref{sec3}. 


\begin{table}[]
\caption{\label{t1} Parameter values and statistical analyses
of the logistic modelling of the World Bank GDP
data~\citep{usgd,cngd,jpgd,degd,ukgd,ingd}
of the six countries that are listed in the first column. The
country-wise ranking is in the order of decreasing GDP
till the year 2020.\footnote{In 2022 India is in the fifth
position and UK is in the sixth.}
The second and third columns list the values of the parameters
$\gamma$ and $k$ for fitting Eq.~(\ref{sologis}) with the GDP
data.%~\citep{usgd,cngd,jpgd,degd,ukgd,ingd}. 
The data have been
plotted and modelled in Figs.~\ref{f1},~\ref{f2}~and~\ref{f3}.
The fourth and fifth columns list, respectively, the mean $\mu$
and the standard deviation $\sigma$ of the yearly {\em relative}
variations of the GDP data
%~\citep{usgd,cngd,jpgd,degd,ukgd,ingd}
with respect to the logistic model.
From the last two columns it is clear that the GDP data and the
logistic model are most closely matched for USA and least
closely matched for China.
}
\begin{ruledtabular}
%\begin{tabular}{|c|ccc|ccc|c|c|}
% &\multicolumn{2}{c}{x}&\multicolumn{2}{c}{y}\\

%\begin{tabular}{|c|c|c|c|c|}
\begin{tabular}{ccccc}

Country
& $\gamma$ & $k$ & $\mu$ & $\sigma$ \\ \hline
%\hline

{USA}
& $0.080$ & $30.0$ & $~~~0.0492$ & $0.0873$ \\

{China}
& $0.095$ & $80.0$ & $-0.3568$ & $0.2504$ \\

{Japan}
& $0.175$ & $~~5.2$ & $-0.0833$ & $0.1395$ \\

{Germany}
& $0.110$ & $~~4.4$ & $~~~0.0489$ & $0.1744$ \\

{UK}
& $0.105$ & $~~3.0$ & $-0.1089$ & $0.1651$ \\

{India}
& $0.080$ & $~~6.0$ & $-0.1359$ & $0.1743$ \\

%\hline
\end{tabular}
\end{ruledtabular}
\end{table}

\section{Forecasting GDP competitiveness}
\label{sec3}
From the GDP values in Figs.~\ref{f1},~\ref{f2}~and~\ref{f3} we 
realize that 
USA and China are at present the top two national economies 
of the world, both with an emphatic lead over the other four 
countries. Although it is unlikely that in the near future the 
GDP of Japan, 
Germany, UK or India may surpass the GDP of either USA or China, 
between USA and China themselves, the GDP gap is reducing 
progressively, as Fig.~\ref{f1} shows. At this rate
the GDP of China may surpass the GDP of USA. 
The year of this overtake can be identified as the 
year when the extrapolated logistic function of the China GDP 
crosses the extrapolated logistic function of the USA GDP. 
However, while the GDP growth of USA is modelled accurately
by the logistic function (a claim supported by the clean fit of
the upper plot in Fig.~\ref{f1}, and the low values 
of $\mu$ and $\sigma$ for USA in Table~\ref{t1}), the same 
observation does not apply to China. The inadequacy of the 
logistic function to model the GDP growth of China is evident 
from the lack of closeness between the logistic function and 
the erratic GDP data in the lower plot in Fig.~\ref{f1}, as 
well as from the high values of $\mu$ and $\sigma$ for 
China in Table~\ref{t1}. That the logistic function falls short 
in modelling the GDP growth of China is known~\citep{kr22}. 
It has been argued that the logistic function properly models 
the GDP growth of countries that foster a democratic polity, 
are free of military conflicts on their borders, and promote free 
economic growth without excessive interference from 
the state~\citep{kr22}. 
The cumulative effect of these conditions is conducive to a 
natural development of material well-being. The absence of 
any one of the aforementioned conditions causes imbalance, 
as happens in the case of China. On the other hand, in the 
other five countries, 
all of the three foregoing conditions prevail in varying degrees, 
and as such, the logistic equation becomes effective in modelling
the GDP growth of these countries~\citep{kr22}. 
This argument is substantiated
by all the related plots in Figs.~\ref{f1},~\ref{f2}~and~\ref{f3}, 
along with the corresponding values of $\mu$ 
and $\sigma$ in Table~\ref{t1}. 

\begin{figure}[]
\begin{center}
\includegraphics[scale=0.65, angle=0]{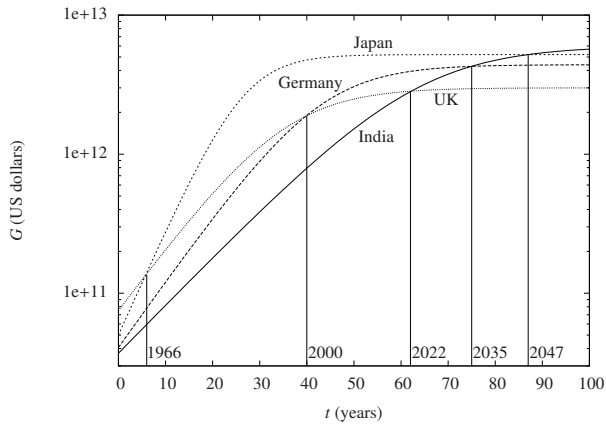}
\caption{\label{f4}\small{
Forecasting the long term outcome of the GDP competitiveness among
Japan, Germany, UK and India. The four theoretical logistic functions,
pertaining to the aforementioned four countries, are calibrated with
the annual GDP data of the World Bank
till 2020~\citep{jpgd,degd,ukgd,ingd}. Two crossings of the theoretical
functions occur before 2020, one in 1966, when Japan overtook UK, and
the other in 2000, when Germany overtook UK. The years of these
intersections are correctly borne out by the World Bank
data~\citep{jpgd,degd,ukgd}. The remaining three intersections
are to occur after 2020, and, hence, are predictive in nature. 
%As it happens, the GDP of India has indeed overtaken the GDP of UK in 2022. 
The first of these, in 2022, has already happened, when the GDP of 
India overtook the GDP of UK. 
Thereafter, the successive overtakes of the 
GDPs of Germany and Japan are predicted to occur in the years 2035
and 2047, respectively. In this plot, the relative positions of
Japan and Germany remain qualitatively unchanged throughout. However,
UK, which began ahead of the other three countries in 1960,
brings up the rear of the group from 2022 onwards. In contrast,
India, which started behind the others in 1960, is to
lead the group from 2047 onwards.
}}
\end{center}
\end{figure}

Considering that China is an anomalous case in modelling 
the dynamics of GDP with the logistic equation, we make 
no further attempt to compare the logistic growth of the GDPs 
of USA and China for predicting the year in which the GDP of the 
latter will overtake the GDP of the former. Instead we 
study the competitiveness of the GDPs of the other five 
countries. However, the GDP of USA is so far ahead 
of the others that in the foreseeable future none of the GDPs  
of Japan, Germany, UK and India is likely to grow close 
enough to the GDP of USA. In that case, a study of the competitiveness 
of GDP growth is meaningful only among Japan, Germany, UK and India. 
Accordingly, it is for these four countries that we 
forecast the outcome of competitive GDP growth. 
Our method consists of extrapolating the theoretical logistic 
functions of Japan, Germany, UK and India beyond the year 2020 
in a single graph, and noting the crossing points among the 
function curves. At the crossing points the GDP of one country 
overtakes the GDP of another. 
Since the four logistic functions have been calibrated with 
the GDP data available up to 2020~\citep{jpgd,degd,ukgd,ingd}, 
any crossing beyond this year enables us to forecast future GDP 
competitiveness among the four countries.  
The result of this whole exercise is to be seen in Fig.~\ref{f4}. 

We first note that Fig.~\ref{f4} has five crossing points. Of these, 
two occur in the years 1966 and 2000,
in both of which the GDP of UK was successively overtaken by the 
GDPs of Japan and Germany. The actual GDP data~\citep{jpgd,degd,ukgd} 
do agree with these intersections, and thus confirm the 
fundamental soundness
of the logistic modelling of GDP growth. While the crossings 
of 1966 and 2000
occur within the range of the available GDP data, i.e. till the year 
2020, there are three more crossing points beyond 2020. These are
in the years 2022, 2035 and 2047, in all of which, the GDP of India
is predicted to successively overtake the GDPs of UK, Germany and 
Japan. %, in that order. 
As it happens, fulfilling the first prediction precisely, 
in the year 2022 the GDP of India has indeed overtaken the GDP of UK. 
%in the year 2022. 
This certainly inspires confidence in the 
predictive power of the logistic modelling of GDP growth. 

Another noteworthy aspect of Fig.~\ref{f4} is that in the year 1960 
(at $t=0$ in the graph) it shows India to have the lowest GDP among 
the four national economies that we compare. The explanation for
this lies in the history of the latter half of the twentieth century. 
In the years following the Second World War, which ended in 1945, 
it became a policy imperative for USA (mainly due to the Cold 
War against the erstwhile Soviet Union) to aid and expedite 
the economic revival of both war-ravaged Japan and Western Europe 
(the latter under the Marshall Plan).  
Guided by USA thus, Japan, Germany (then West Germany) and UK
had achieved political peace and economic prosperity by 1960. 
In contrast, during the same period, India, freed from 
colonial rule about a decade earlier, did not experience the 
advantages that regenerated the 
economies of Japan, Germany and UK. Two factors, more than any other,  
impeded the GDP growth of India. The first is government policies in 
economic matters, and the second is a series of wars %military conflicts 
in which India was embroiled
in the initial three decades of its sovereign existence. 
Unsurprisingly then, the GDP growth of India is seen to trail
those of the other countries in Fig.~\ref{f4} from 1960 to 2020. 
And yet by 2047, the GDP of India is projected in Fig.~\ref{f4} 
to lead the GDPs of the other three countries. 
This will be possible only because India
has maintained a steady GDP growth rate over an extended duration. 
One reason for this sustained growth rate is that India is a country 
of sub-continental proportions with a large population, unlike Japan, 
Germany and UK. Now, it is known that the GDP of a national economy 
is scaled as a function of its trade by a power law, 
%\begin{equation} 
%\label{scale} 
%G(T) \sim T^{\alpha},
%\end{equation}
$G \sim T^\alpha$, 
in which $T$ is the trade 
volume and $\alpha \,(>0)$ is the power-law exponent~\citep{kr22}.
For all the countries that we study here, the power-law
scaling is known to hold true over at least
two orders of magnitude~\citep{kr22}. What is more, the 
exponent $\alpha$ distinguishes the economies of large 
countries (with large areas and populations) from the economies of 
small ones (with small areas and populations)~\citep{kr22}. 
In the former type, which includes India (as well as USA and China), 
$\alpha$ has a relatively 
low value~\citep{kr22}. In the latter type, which includes Japan, 
Germany and UK, $\alpha$ has a higher value~\citep{kr22}.

We now explain how the distinction between the two types of national 
economies can cause
%qualitative 
a difference in trading patterns, with a 
concomitant effect on the GDP growth. A country with a large population 
has the advantage of a proportionately large domestic consumption
of its own products, which in turn makes a 
proportionately greater 
contribution to the GDP, as compared to the contribution from trade. 
Consequently, the %percentage 
contribution of trade to the overall GDP growth reduces, a condition 
that is reflected by a lower value of $\alpha$. %in Eq.~\ref{scale}. 
The same feature is also 
known for USA and China, which like India, are geographically 
extended countries with large populations~\citep{kr22}. Countries with 
small populations, on the other hand, are bereft of the means 
of the high 
domestic consumption that attends a large population, and 
thus they have to rely more on trade with other countries to enhance 
their GDPs~\citep{fr99}. 
This condition is reflected by a higher 
value of $\alpha$, as is known to happen in the case of Japan, 
Germany and UK~\citep{kr22}. Now, the growth of trade also saturates 
according to the logistic function~\citep{kr22}, and since trade
is highly correlated with GDP~\citep{kr22}, a saturation of trade 
implies a corresponding saturation of the GDP. Therefore, countries 
that depend more on external trade than on domestic consumption will 
see their GDP growth saturate when their trade saturates due to 
market-driven inhibitors. This is what 
we see for Japan, Germany and UK in Fig.~\ref{f4}, whereas the 
GDP of India, despite its slow start, will outpace the GDP of 
the other three countries on the strength of its large volume 
of domestic consumption. 

\section{Conclusions and remarks}
\label{sec4} 
The suitability of the logistic equation to model the dynamics
of GDP and trade has been established already~\citep{kr22}. 
In the present study we proceed further to show that the logistic
equation is also effective in forecasting the outcome of 
GDP competitiveness among some leading national economies. 
Our logistic forecasting method has been vindicated both 
retrospectively and for future times. In the former case, it 
has correctly estimated the years when the GDPs 
of Japan and Germany overtook the GDP of UK (1966 and 2000, 
respectively). In the latter case, looking forward in time, the logistic 
method has also been successful in forecasting 2022 as the year
in which the GDP of India is to overtake the GDP of UK. What 
now remains to be seen is the fulfillment of the forecast that 
the GDP of India will overtake the GDPs of Germany and Japan 
in 2035 and 2047, respectively. 
  
All that said, we have to remember that the subject of our 
present study is the dynamics of social systems (national 
economies). Hence, it must depend on real socio-economic data. 
For instance, the logistic functions in Fig.~\ref{f4},
which are at the core of our forecasts, have all been calibrated    
with the GDP data of the World Bank till
2020~\citep{usgd,cngd,jpgd,degd,ukgd,ingd}.
However, unforeseen
natural, social and political events can compromise 
our forecasts by recalibrating the parameters of the
logistic equation. To understand what such events may be like, 
we consider some contemporary examples.  
Even in the first quarter of 2020 no one 
anticipated that within two years the global 
economy was to receive three major shocks. These are, first, 
Covid-19, which became a global pandemic by the middle of 2020
and whose aftereffects are felt even in late 2022. The second 
shock has been  
the war in Ukraine, which broke out in early 2022, even before the world 
economy could fully recover from the damage it had suffered due to 
the Covid-19 
pandemic. The Ukraine war, by now protracted beyond all 
expectations, has disrupted 
global supply chains and energy markets, the severity of which 
is not yet fully understood. The third shock has 
been droughts in Europe and China, which will
affect crucial sectors like power and agriculture. Considering that 
Europe and China belong to two major 
economic regions of the world, the former in the North-Atlantic
and the latter in the Indo-Pacific, adverse climatic events in these 
regions will have an adverse impact on production globally. 
 
%\begin{acknowledgments}
%The author thanks A. Kakkad. 
% and N. Sarkar. Comments from
%J. Mulherkar, A. Parikh and M. Tiwari are appreciated.  
%\end{acknowledgments}

\bibliography{arXgdp22}

\begin{thebibliography}{17}%
\makeatletter
\providecommand \@ifxundefined [1]{%
 \@ifx{#1\undefined}
}%
\providecommand \@ifnum [1]{%
 \ifnum #1\expandafter \@firstoftwo
 \else \expandafter \@secondoftwo
 \fi
}%
\providecommand \@ifx [1]{%
 \ifx #1\expandafter \@firstoftwo
 \else \expandafter \@secondoftwo
 \fi
}%
\providecommand \natexlab [1]{#1}%
\providecommand \enquote  [1]{``#1''}%
\providecommand \bibnamefont  [1]{#1}%
\providecommand \bibfnamefont [1]{#1}%
\providecommand \citenamefont [1]{#1}%
\providecommand \href@noop [0]{\@secondoftwo}%
\providecommand \href [0]{\begingroup \@sanitize@url \@href}%
\providecommand \@href[1]{\@@startlink{#1}\@@href}%
\providecommand \@@href[1]{\endgroup#1\@@endlink}%
\providecommand \@sanitize@url [0]{\catcode `\\12\catcode `\$12\catcode
  `\&12\catcode `\#12\catcode `\^12\catcode `\_12\catcode `\%12\relax}%
\providecommand \@@startlink[1]{}%
\providecommand \@@endlink[0]{}%
\providecommand \url  [0]{\begingroup\@sanitize@url \@url }%
\providecommand \@url [1]{\endgroup\@href {#1}{\urlprefix }}%
\providecommand \urlprefix  [0]{URL }%
\providecommand \Eprint [0]{\href }%
\providecommand \doibase [0]{http://dx.doi.org/}%
\providecommand \selectlanguage [0]{\@gobble}%
\providecommand \bibinfo  [0]{\@secondoftwo}%
\providecommand \bibfield  [0]{\@secondoftwo}%
\providecommand \translation [1]{[#1]}%
\providecommand \BibitemOpen [0]{}%
\providecommand \bibitemStop [0]{}%
\providecommand \bibitemNoStop [0]{.\EOS\space}%
\providecommand \EOS [0]{\spacefactor3000\relax}%
\providecommand \BibitemShut  [1]{\csname bibitem#1\endcsname}%
\let\auto@bib@innerbib\@empty
%</preamble>
\bibitem [{\citenamefont {Strogatz}(1994)}]{stro}%
  \BibitemOpen
  \bibfield  {author} {\bibinfo {author} {\bibfnamefont {S.~H.}\ \bibnamefont
  {Strogatz}},\ }\href@noop {} {\emph {\bibinfo {title} {Nonlinear Dynamics and
  Chaos}}}\ (\bibinfo  {publisher} {Addison-Wesley Publishing Company},\
  \bibinfo {address} {Reading, MA},\ \bibinfo {year} {1994})\BibitemShut
  {NoStop}%
\bibitem [{\citenamefont {Braun}(1983)}]{braun}%
  \BibitemOpen
  \bibfield  {author} {\bibinfo {author} {\bibfnamefont {M.}~\bibnamefont
  {Braun}},\ }\href@noop {} {\emph {\bibinfo {title} {Differential Equations
  and Their Applications}}}\ (\bibinfo  {publisher} {Springer-Verlag},\
  \bibinfo {address} {New York},\ \bibinfo {year} {1983})\BibitemShut {NoStop}%
\bibitem [{\citenamefont {Montroll}(1978)}]{mon78}%
  \BibitemOpen
  \bibfield  {author} {\bibinfo {author} {\bibfnamefont {E.~W.}\ \bibnamefont
  {Montroll}},\ }\href@noop {} {\bibfield  {journal} {\bibinfo  {journal}
  {Proc. Natl. Acad. Sci. USA}\ }\textbf {\bibinfo {volume} {75}},\ \bibinfo
  {pages} {4633} (\bibinfo {year} {1978})}\BibitemShut {NoStop}%
\bibitem [{\citenamefont {Ray}(2010)}]{akr10}%
  \BibitemOpen
  \bibfield  {author} {\bibinfo {author} {\bibfnamefont {A.~K.}\ \bibnamefont
  {Ray}},\ }\bibfield  {title} {\enquote {\bibinfo {title} {{Modelling
  Saturation in Industrial Growth}},}\ }in\ \href@noop {} {\emph {\bibinfo
  {booktitle} {Econophysics \& Economics of Games, Social Choices and
  Quantitative Techniques}}},\ \bibinfo {editor} {edited by\ \bibinfo {editor}
  {\bibfnamefont {B.}~\bibnamefont {Basu}}, \bibinfo {editor} {\bibfnamefont
  {B.~K.}\ \bibnamefont {Chakrabarti}}, \bibinfo {editor} {\bibfnamefont
  {S.~R.}\ \bibnamefont {Chakravarty}}, \ and\ \bibinfo {editor} {\bibfnamefont
  {K.}~\bibnamefont {Gangopadhyay}}}\ (\bibinfo  {publisher} {Springer-Verlag
  Italia},\ \bibinfo {address} {Milan},\ \bibinfo {year} {2010})\BibitemShut
  {NoStop}%
\bibitem [{\citenamefont {Kakkad}\ and\ \citenamefont {Ray}(2023)}]{kr22}%
  \BibitemOpen
  \bibfield  {author} {\bibinfo {author} {\bibfnamefont {A.}~\bibnamefont
  {Kakkad}}\ and\ \bibinfo {author} {\bibfnamefont {A.~K.}\ \bibnamefont
  {Ray}},\ }\href@noop {} {\bibfield  {journal} {\bibinfo  {journal} {Int. J.
  Mod. Phys. C}\ }\textbf {\bibinfo {volume} {(Online Ready)}},\ \bibinfo
  {pages} {https://doi.org/10.1142/S0129183123500201} (\bibinfo {year}
  {2023})}\BibitemShut {NoStop}%
\bibitem [{\citenamefont {Samuelson}\ and\ \citenamefont
  {Nordhaus}(1998)}]{samnord}%
  \BibitemOpen
  \bibfield  {author} {\bibinfo {author} {\bibfnamefont {P.~A.}\ \bibnamefont
  {Samuelson}}\ and\ \bibinfo {author} {\bibfnamefont {W.~D.}\ \bibnamefont
  {Nordhaus}},\ }\href@noop {} {\emph {\bibinfo {title} {Economics}}}\
  (\bibinfo  {publisher} {Tata McGraw-Hill},\ \bibinfo {address} {New Delhi},\
  \bibinfo {year} {1998})\BibitemShut {NoStop}%
\bibitem [{\citenamefont {Serrano}(2007)}]{mas07}%
  \BibitemOpen
  \bibfield  {author} {\bibinfo {author} {\bibfnamefont {M.~{\'A}ngeles}\
  \bibnamefont {Serrano}},\ }\href@noop {} {\bibfield  {journal} {\bibinfo
  {journal} {J. Stat. Mech.}\ }\textbf {\bibinfo {volume} {01}},\ \bibinfo
  {pages} {L01002} (\bibinfo {year} {2007})}\BibitemShut {NoStop}%
\bibitem [{\citenamefont {Garlaschelli}\ \emph {et~al.}(2007)\citenamefont
  {Garlaschelli}, \citenamefont {Matteo}, \citenamefont {Aste}, \citenamefont
  {Caldarelli},\ and\ \citenamefont {Loffredo}}]{gmacl07}%
  \BibitemOpen
  \bibfield  {author} {\bibinfo {author} {\bibfnamefont {D.}~\bibnamefont
  {Garlaschelli}}, \bibinfo {author} {\bibfnamefont {T.~Di}\ \bibnamefont
  {Matteo}}, \bibinfo {author} {\bibfnamefont {T.}~\bibnamefont {Aste}},
  \bibinfo {author} {\bibfnamefont {G.}~\bibnamefont {Caldarelli}}, \ and\
  \bibinfo {author} {\bibfnamefont {M.~I.}\ \bibnamefont {Loffredo}},\
  }\href@noop {} {\bibfield  {journal} {\bibinfo  {journal} {Eur. Phys. J. B}\
  }\textbf {\bibinfo {volume} {57}},\ \bibinfo {pages} {159} (\bibinfo {year}
  {2007})}\BibitemShut {NoStop}%
\bibitem [{\citenamefont {Garlaschelli}\ and\ \citenamefont
  {Loffredo}(2004)}]{gl04}%
  \BibitemOpen
  \bibfield  {author} {\bibinfo {author} {\bibfnamefont {D.}~\bibnamefont
  {Garlaschelli}}\ and\ \bibinfo {author} {\bibfnamefont {M.~I.}\ \bibnamefont
  {Loffredo}},\ }\href@noop {} {\bibfield  {journal} {\bibinfo  {journal}
  {Phys. Rev. Lett.}\ }\textbf {\bibinfo {volume} {93}},\ \bibinfo {pages}
  {188701} (\bibinfo {year} {2004})}\BibitemShut {NoStop}%
\bibitem [{\citenamefont {Albert}\ and\ \citenamefont
  {Barab\'asi}(2002)}]{albar02}%
  \BibitemOpen
  \bibfield  {author} {\bibinfo {author} {\bibfnamefont {R.}~\bibnamefont
  {Albert}}\ and\ \bibinfo {author} {\bibfnamefont {{A.-L.}}\ \bibnamefont
  {Barab\'asi}},\ }\href@noop {} {\bibfield  {journal} {\bibinfo  {journal}
  {Rev. Mod. Phys.}\ }\textbf {\bibinfo {volume} {74}},\ \bibinfo {pages} {47}
  (\bibinfo {year} {2002})}\BibitemShut {NoStop}%
\bibitem [{usg()}]{usgd}%
  \BibitemOpen
  \href@noop {} {\enquote {\bibinfo {title} {{USA GDP data}},}\ }\bibinfo
  {howpublished} {\url{
  https://data.worldbank.org/indicator/NY.GDP.MKTP.CD?locations=US
  }}\BibitemShut {NoStop}%
\bibitem [{cng()}]{cngd}%
  \BibitemOpen
  \href@noop {} {\enquote {\bibinfo {title} {{China GDP data}},}\ }\bibinfo
  {howpublished} {\url{
  https://data.worldbank.org/indicator/NY.GDP.MKTP.CD?locations=CN
  }}\BibitemShut {NoStop}%
\bibitem [{jpg()}]{jpgd}%
  \BibitemOpen
  \href@noop {} {\enquote {\bibinfo {title} {{Japan GDP data}},}\ }\bibinfo
  {howpublished} {\url{
  https://data.worldbank.org/indicator/NY.GDP.MKTP.CD?locations=JP
  }}\BibitemShut {NoStop}%
\bibitem [{deg()}]{degd}%
  \BibitemOpen
  \href@noop {} {\enquote {\bibinfo {title} {{Germany GDP data}},}\ }\bibinfo
  {howpublished} {\url{
  https://data.worldbank.org/indicator/NY.GDP.MKTP.CD?locations=DE
  }}\BibitemShut {NoStop}%
\bibitem [{ukg()}]{ukgd}%
  \BibitemOpen
  \href@noop {} {\enquote {\bibinfo {title} {{UK GDP data}},}\ }\bibinfo
  {howpublished} {\url{
  https://data.worldbank.org/indicator/NY.GDP.MKTP.CD?locations=GB
  }}\BibitemShut {NoStop}%
\bibitem [{ing()}]{ingd}%
  \BibitemOpen
  \href@noop {} {\enquote {\bibinfo {title} {{India GDP data}},}\ }\bibinfo
  {howpublished} {\url{
  https://data.worldbank.org/indicator/NY.GDP.MKTP.CD?locations=IN
  }}\BibitemShut {NoStop}%
\bibitem [{\citenamefont {Frankel}\ and\ \citenamefont {Romer}(1999)}]{fr99}%
  \BibitemOpen
  \bibfield  {author} {\bibinfo {author} {\bibfnamefont {J.~A.}\ \bibnamefont
  {Frankel}}\ and\ \bibinfo {author} {\bibfnamefont {D.}~\bibnamefont
  {Romer}},\ }\href@noop {} {\bibfield  {journal} {\bibinfo  {journal} {The
  American Economic Review}\ }\textbf {\bibinfo {volume} {89}},\ \bibinfo
  {pages} {379} (\bibinfo {year} {1999})}\BibitemShut {NoStop}%
\end{thebibliography}%
%\printbibliography
\end{document}